\documentclass[a4paper,12pt]{article}
\usepackage[dvips]{graphicx}
\usepackage{graphics}

\title{Electromagnetic diffraction on a moving half-plane}

\author{A.~Ciarkowski and B.~Atamaniuk\\
\normalsize Institute of Fundamental Technological Research\\
\normalsize Polish Academy of Sciences}

\date{}

\newcommand{\p}{^\prime}
\renewcommand{\b}{\beta}
\renewcommand{\o}{\omega}
\renewcommand{\t}{\theta}
\newcommand{\g}{\gamma}
\renewcommand{\r}{\rho}
\newcommand{\f}{\phi}
\newcommand{\mb}[1]{\mbox{\boldmath $#1$}}

\begin{document}

\maketitle

\abstract{A problem of electromagnetic (EM) plane wave diffraction
on a moving half-plane in a homogeneous and isotropic medium is
considered. It is shown, that unlike the stationary case, the
shadow boundaries of the incident and reflected wave are not
parallel to propagation directions of those waves. Other
diffraction phenomena reported earlier for objects with moving
edges, are also observed here.}

\section{Introduction}

Studying wave scattering by objects in motion has a long history.
This sort of EM field interaction with obstacles is characteristic
of object recognition, space science and astronomy. Objects with
sharp edges form a subclass of obstacles which are of both
theoretical and practical interest. Phenomena associated with
electromagnetic scattering on moving objects with edges, like
Doppler shift of equiphase surfaces in the diffracted wave and
angular shift of the location of its amplitude singularities, were
observed in \cite{lm;67} and \cite{gt;68}. In more recent works
(\cite{dc;00}, \cite{dc;01}, \cite{dc;02}, \cite{dc;03}) new
analytic and numerical results were obtained, which include
alternative representation for the moving wedge solution,
development of new numerical techniques, and extension of
interacting fields to Gaussian beams.

A special problem in this area is that of EM diffraction by a
moving half-plane. This problem is specially convenient for
studying phenomena characteristic of a moving edge, because the
location of shadow boundaries, and thus location of lit and shadow
regions of geometrical optics (GO) field, can be inferred directly
from the arguments of special functions defining the solution.
Extra information on diffraction phenomena can be gained from
studying high-frequency asymptotic expansion of the solution.

In this paper we assume that the field incident on the moving
half-plane is a plane wave, and that the incidence direction is
perpendicular to the half-plane edge. With the help of Lorentz
transformation and the use of well known Sommerfeld solution
obtained in stationary case the total field is found. This field
is then asymptotically expanded and interpreted physically in
terms of generated waves.

It is shown, that in addition to phenomena observed in the
diffracted wave (comp.\ \cite{gt;68}), domains of existence of
geometrical optics field are also modified.

\section{Formulation of the problem}
Assume that a perfectly conducting half-plane is in steady motion
in a homogeneous and isotropic medium. We introduce two coordinate
systems $(x,y,z,t)$ and $(x\p,y\p,z\p,t\p)$, the first associated
with the laboratory frame, and the other with the moving
half-plane. To an observer in the primed system the half-plane is
stationary and is described by $x\p\le 0$, $-\infty<y\p<\infty$
and $z\p=0$. To an observer in the laboratory system, the primed
system progresses along the positive $x-$axis with the constant
velocity \mb v = \mb {\hat{x}}$v$. The two other axes, $y\p$ and
$z\p$, displace in space parallel to the corresponding axes $y$
and $z$ with the same velocity $v$.

Assume further that in the laboratory system, EM plane wave,
varying in time as $\exp{(-i\o t)}$, propagates in the direction
perpendicular to the half-plane edge. Thus the problem is
2-dimensional and any EM field containing both electric and
magnetic $y$-components can be decomposed into \textit{E-} and
\textit{H-}parts with respect to $y$-axis, where those components
are separated (\cite{bw;64}). In particular, a propagating plane
wave can be represented as a sum of \emph{E}-wave,
\begin{equation}\label{e1}
\mb E^i = [0, A_1,0] e^{-ik(x\cos\t + z\sin\t + ct)}, \quad  c\mb
B^i = \frac{1}{ik} \nabla\times\mb E^i,
\end{equation}
and \emph{H}-wave,
\begin{equation}\label{e2}
c\mb B^i = [0, A_2,0] e^{-ik(x\cos\t + z\sin\t + ct)}, \quad  \mb
E^i = -\frac{1}{ik} \nabla\times c \mb B.
\end{equation}
Here, $k$ and $c$ are the wave number and the speed of light,
respectively, $\t\in(-\pi,\pi)$ is the incident angle measured
from the positive $x-$axis to the direction from which the wave
propagates, and $A_i$, $i=1,2$, are the amplitudes of the $y$
components of the respective plane waves. (The ratio $A_1/A_2$
determines the polarization of the wave).

Thus a general, $y$-independent incident wave can be written down
as
\begin{equation}\label{e3}
\begin{array}{r}
\displaystyle \mb E^i = [-A_2\sin\t, A_1,A_2\cos\t] e^{-ik(x\cos\t
+
z\sin\t + ct)}, \\[5pt] \displaystyle c\mb B^i =
[A_1\sin\t,A_2,-A_1\cos\t] e^{-ik(x\cos\t + z\sin\t + ct)},
\end{array}
\end{equation}
which is a sum of two plane waves of both types.

In what follows we find the field resulting from scattering of
this wave on the moving half-plane and emphasize differences in
its structure in comparison to the stationary case.

\section{The scattered field}
Our construction of the scattered field reduces to Lorentz
transformation of the incident wave from the laboratory to the
moving frame, employing there the well known Sommerfeld solution
for half-plane diffraction, and finally transformation of that
solution back to the laboratory frame.

\subsection{Lorentz transformation of the incident wave to the moving
frame} Lorentz transformation of (\ref{e3}) leads to
\begin{equation}\label{e4}
\begin{array}{r}
\displaystyle \mb {E^i}\p = [-A_2\p\sin\t\p, A_1\p,A_2\p\cos\t\p]
e^{-ik\p(x\p\cos\t \p+
z\p\sin\t\p + ct\p)}, \\[5pt] \displaystyle c\mb {B^i}\p =
[A_1\p\sin\t\p,A_2\p,-A_1\p\cos\t\p] e^{-ik\p(x\p\cos\t\p +
z\p\sin\t\p + ct\p)},
\end{array}
\end{equation}
where
\[\g=(1-\b^2)^{-1/2}, \qquad \b=\frac{v}{c}, \]
%Comparison of (\ref{e4}) and (\ref{e9}) shows that
\begin{equation}\label{e10}
A_i\p=A_i \g (1+\b\cos\t), \quad i=1,2,
\end{equation}
\begin{equation}\label{e11}
\cos\t\p=\frac{\cos\t+\b}{1+\b\cos\t}\;, \quad
\sin\t\p=\frac{\sin\t}{\g(1+\b\cos\t)}\:,
\end{equation}
and
\begin{equation}\label{e11a}
k\p=k\g(1+\b\cos\t).
\end{equation}
The last formula is a manifestation of Doppler effect: if the
plane wave propagates in the direction opposite to \mb v, then
$k\p=k\sqrt{(1+\b)/(1-\b)}>k$, and hence $\o\p>\o$. In this case
the frequency in the moving frame is higher than that in the
laboratory frame. Conversely, if the wave progresses in the
direction coinciding with \mb v, then
$k\p=k\sqrt{(1-\b)/(1+\b)}<k$, and $\o\p<\o$.

If the incident wave (\ref{e3}) propagates perpendicularly  to \mb
v, i.e.\ if $\t=\pm\pi/2$, then by (\ref{e11}),
$\t\p=\pm(\pi/2-\arcsin{\b})$ and the direction of wave
propagation in the primed system has a tendency to rotate in the
direction opposite to that of \mb v.

\subsection{The total field in the moving frame}
The total field in the moving frame is obtained by the use of the
well known Sommerfeld solution for stationary half-plane
\cite{no;58}, together with the transformed incident wave
(\ref{e4}). The corresponding solutions for each field type are as
follows:
\subsubsection{\textit{E-}field}
In this case \mb E$\p$ = \mb {\hat{y}\p}$E\p_{y\p}$, and the only
component of the electric field is
\begin{equation}\label{e12}
E\p_{y\p} = A_1\p \frac{e^{-i\pi/4}}{\sqrt{\pi}}e^{ik\p \r\p}
[G(u\p) - G(v\p)]e^{-ik\p ct\p}.
\end{equation}
Here,
\begin{equation}\label{e13}
G(a) = e^{-ia^2}\int_a^\infty e^{i\tau^2}\;d\tau,
\end{equation}
\begin{equation}\label{e14}
u\p=-\sqrt{2 k\p \r\p} \cos{\frac{\f\p-\t\p}{2}}, \qquad
v\p=\sqrt{2 k\p \r\p} \cos{\frac{\f\p+\t\p}{2}},
\end{equation}
$\r\p$ and $\t\p$ are cylindrical coordinates in the primed frame
and the incident angle $\t\p$ is defined by (\ref{e11}).

The components of the magnetic induction are by (\ref{e1}) found
to be
\begin{equation}\label{e15}
cB\p_{x\p} = A_1\p \frac{e^{-i\pi/4}}{\sqrt{\pi}}e^{ik\p \r\p}
\left\{\sin{\t\p}[G(u\p) + G(v\p)] + i\sqrt{\frac{2}{k\p
\r\p}}\sin{\frac{\f\p}{2}}\cos{\frac{\t\p}{2}} \right\}e^{-ik\p
ct\p}
\end{equation}
and
\begin{equation}\label{e16}
cB\p_{z\p} = A_1\p \frac{e^{-i\pi/4}}{\sqrt{\pi}}e^{ik\p \r\p}
\left\{\cos{\t\p}[G(u\p) - G(v\p)] - i\sqrt{\frac{2}{k\p
\r\p}}\cos{\frac{\f\p}{2}}\cos{\frac{\t\p}{2}} \right\}e^{-ik\p
ct\p}.
\end{equation}

\subsubsection{\textit{H-}field}
Here, $c$\mb B$\p$ = \mb {\hat{y}\p}$cB\p_{y\p}$, where
\begin{equation}\label{e17}
cB\p_{y\p} = A_2\p \frac{e^{-i\pi/4}}{\sqrt{\pi}}e^{ik\p \r\p}
[G(u\p) + G(v\p)]e^{-ik\p ct\p}.
\end{equation}
The electric field is given by (comp.\ (\ref{e2}))
\begin{equation}\label{e18}
E\p_{x\p} = A_2\p \frac{e^{-i\pi/4}}{\sqrt{\pi}}e^{ik\p \r\p}
\left\{\sin{\t\p}[G(u\p) - G(v\p)] - i\sqrt{\frac{2}{k\p
\r\p}}\cos{\frac{\f\p}{2}}\sin{\frac{\t\p}{2}} \right\}e^{-ik\p
ct\p}
\end{equation}
and
\begin{equation}\label{e19}
E\p_{z\p} = A_2\p \frac{e^{-i\pi/4}}{\sqrt{\pi}}e^{ik\p \r\p}
\left\{\cos{\t\p}[G(u\p) + G(v\p)] - i\sqrt{\frac{2}{k\p
\r\p}}\sin{\frac{\f\p}{2}}\sin{\frac{\t\p}{2}} \right\}e^{-ik\p
ct\p}.
\end{equation}

\subsection{The total field in the laboratory frame}
Transformation of these fields back to the laboratory frame leads
to the following expressions for both polarizations in the
laboratory frame:

\subsubsection{\textit{E-}field}
\begin{equation}\label{e22}
\mb E_\bot = A_1\p\g\; \mb {\hat{y}}\;
\frac{e^{-i\pi/4}}{\sqrt{\pi}}e^{ik\p \r\p}\left\{(1-\b\cos{\t\p})
[G(u\p) - G(v\p)]- i\sqrt{\frac{2}{k\p
\r\p}}\cos{\frac{\f\p}{2}}\cos{\frac{\t\p}{2}} \right\} e^{-ik\p
ct\p},
\end{equation}
\begin{equation}\label{e23}
c\mb B_\| = A_1\p\; \mb {\hat{x}}\;
\frac{e^{-i\pi/4}}{\sqrt{\pi}}e^{ik\p \r\p}\left\{\sin{\t\p}
[G(u\p) + G(v\p)]+ i\sqrt{\frac{2}{k\p
\r\p}}\sin{\frac{\f\p}{2}}\cos{\frac{\t\p}{2}} \right\} e^{-ik\p
ct\p},
\end{equation}
\begin{equation}\label{e24}
c\mb B_\bot = A_1\p\g\; \mb {\hat{z}}\;
\frac{e^{-i\pi/4}}{\sqrt{\pi}}e^{ik\p \r\p}\left\{(\b-\cos{\t\p})
[G(u\p) - G(v\p)]- i\sqrt{\frac{2}{k\p
\r\p}}\cos{\frac{\f\p}{2}}\cos{\frac{\t\p}{2}} \right\} e^{-ik\p
ct\p},
\end{equation}

\subsubsection{\textit{H-}field}
\begin{equation}\label{e25}
c\mb B_\bot = A_2\p\g\; \mb {\hat{y}}\;
\frac{e^{-i\pi/4}}{\sqrt{\pi}}e^{ik\p \r\p}\left\{(1-\b\cos{\t\p})
[G(u\p) + G(v\p)]+ i\sqrt{\frac{2}{k\p
\r\p}}\sin{\frac{\f\p}{2}}\sin{\frac{\t\p}{2}} \right\} e^{-ik\p
ct\p},
\end{equation}
\begin{equation}\label{e26}
\mb E_\| = A_2\p\; \mb {\hat{x}}\;
\frac{e^{-i\pi/4}}{\sqrt{\pi}}e^{ik\p \r\p}\left\{\sin{\t\p}
[G(u\p) - G(v\p)]- i\sqrt{\frac{2}{k\p
\r\p}}\cos{\frac{\f\p}{2}}\sin{\frac{\t\p}{2}} \right\} e^{-ik\p
ct\p},
\end{equation}
\begin{equation}\label{e27}
\mb E_\bot = A_2\p\g\; \mb {\hat{z}}\;
\frac{e^{-i\pi/4}}{\sqrt{\pi}}e^{ik\p \r\p}\left\{(\cos{\t\p}-\b)
[G(u\p) + G(v\p)]- i\sqrt{\frac{2}{k\p
\r\p}}\sin{\frac{\f\p}{2}}\sin{\frac{\t\p}{2}} \right\} e^{-ik\p
ct\p}.
\end{equation}

The total field is the sum of the component fields. For
compactness, these fields are here given in terms of primed
quantities.

\subsection{Asymptotic representation of the total field}
In an attempt to get insight into physical structure of the total
field we asymptotically evaluate it as $k\p$, or frequency, tends
to infinity. The procedure reduces to the asymptotic expansion of
the function $G(\cdot)$ in (\ref{e22}) through (\ref{e27}) as its
argument tends to infinity. The result is (\cite{bw;64})
\begin{equation}\label{e28}
G(a)=H(-a)\sqrt{\pi}\;e^{i\frac{\pi}{4}}\;e^{-ia^2} + \frac{i}{2a
} + O\left(\frac{1}{a^2}\right),
\end{equation}
where $H(\cdot)$ is a Heaviside unit step function.

The term proportional to $H(-a)$ on the rhs of (\ref{e28}) is
responsible for a geometrical optics field contribution, while the
remainder contributes to the diffracted field. The lit and shadow
domains of corresponding geometrical optics fields (here the
incident and reflected waves) are specified by the sign of $a$,
minus sign applies to the lit region, and plus sign to the shadow
region. Those regions are separated by the shadow boundary,
defined by $a=0$.

Application of (\ref{e28}) to (\ref{e22}) through (\ref{e27})
yields the following asymptotic representation of the \textit{E-}
and \textit{H-}field contributions to the total field in the
laboratory frame:
\newline\textsf{\textit{E-}field}:
\begin{equation}\label{e29}
\begin{array}{ll}
\mb E_\bot \sim A_1 \mb {\hat{y}} (u^{GO-}+u^d_{e1}) &  \\[2ex]
c\mb B_\| \sim A_1 \mb {\hat{x}} (\sin\t\; u^{GO+}+u^d_{e2}), &
c\mb B_\bot\sim A_1 \mb {\hat{z}} (-\cos\t\;u^{GO-}+u^d_{e3}),
\end{array}
\end{equation}
\textsf{\textit{H-}field}:
\begin{equation}\label{e30}
\begin{array}{ll}
c\mb B_\bot \sim A_2 \mb {\hat{y}} (u^{GO+}+u^d_{h1}) &  \\[2ex]
\mb E_\| \sim A_2 \mb {\hat{x}} (-\sin\t\;u^{GO-}+u^d_{h2}), & \mb
E_\bot \sim A_2 \mb {\hat{z}} (\cos\t\;u^{GO+}+u^d_{h3}).
\end{array}
\end{equation}
Here,
\begin{equation}\label{e31}
u^{GO\pm}=H(\epsilon^{i\prime})e^{-ik(x\cos{\t}+z\sin{\t})-i\o t}
\pm H(\epsilon^{r\prime})e^{-ik(x\cos{\t}-z\sin{\t})-i\o t}
\end{equation}
is a combination of the incident and the reflected plane wave,
both waves vanishing in their shadow regions. The quantities
\begin{equation}\label{e32}
\begin{array}{lll}
u^d_{e1}=-\g f^+g^+h_1s & u^d_{e2}=\mbox{sign}(z)f^+g^-h^+_2s
& u^d_{e3}=-\g f^+g^+h_3s \\[1.5ex]
u^d_{h1}=\mbox{sign}(\t z) \g f^-g^-h_1s &
u^d_{h2}=\mbox{sign}(\t) f^-g^+h^-_2s & u^d_{h3}=-\mbox{sign}(\t
z) \g f^-g^-h_3s
\end{array}
\end{equation}
represent the Cartesian components of the diffracted waves. The
factors appearing in (\ref{e32}) are given by
\begin{equation}\label{e33}
f^\pm=\sqrt{\frac{(1\pm\b)(1\pm\cos{\t})}{2ik}},
\end{equation}
\begin{equation}\label{e34}
g^\pm=\frac{\sqrt{r\pm(x-\b ct)}}{r}
\frac{1+\b\cos{\t}}{(1+\b\cos{\t})(x-\b ct)+(\cos{\t}+\b)r},
\end{equation}
\begin{equation}\label{e35}
h_1=r+\b(x-\b ct) \qquad h_2^\pm=r\pm(x-\b ct) \qquad h_3=\b
r+x-\b ct,
\end{equation}
\begin{equation}\label{e36}
r=\sqrt{(x-\b ct)^2 + (z/\g)^2},
\end{equation}
and
\begin{equation}\label{e37}
s=\exp{[ik\g^2(1+\b\cos{\t})(r+\b x-ct)-i\pi/4]}.
\end{equation}
The shadow indicators $\epsilon^{i(r)\prime}$ are discussed in the
next section.

\section{Phenomena specific to diffraction on a moving edge}

Essential features in the field resulting from plane wave
scattering by the moving half-plane are similar as in the
stationary case, i.e.\ there are three different wave species
appears in the solution: two plane waves -- incident and reflected
one, and the diffracted wave. Nevertheless, there are phenomena
specific for moving edge diffraction. They are discussed below.

The shadow boundaries in the moving frame are defined by $u\p=0$
and $v\p=0$ for any $\r\p$, or equivalently by
\begin{equation}\label{e38}
\cos{(\f\p \mp \t\p)}=-1, \qquad
\mbox{sign}(x\p)=-\mbox{sign}(\cos{\t\p}).
\end{equation}
Here the upper (lower) sign applies to the incident (reflected)
wave. By multiplying this by $\r\p$, squaring both sides and
solving the resultant equation we find
\begin{equation}\label{e39}
\frac{z\p}{x\p}=\pm\tan{\t\p}, \qquad
\mbox{sign}(x\p)=-\mbox{sign}(\cos{\t\p}).
\end{equation}
This equation transforms in the laboratory frame into
\begin{equation}\label{e40}
\frac{z}{x-vt}=\pm\tan{\t}\;\frac{1}{1+\frac{\b}{\cos{\t}}}\:,
\qquad \mbox{sign}(x-vt)=-\mbox{sign}(\cos{\t}+\b).
\end{equation}
For any time instant $t$, equations (\ref{e40}) describe
semi-infinite straight lines in the ($x$,$z$) plane, with their
starting points at the screen edge. If $0<|\t|<\pi/2$ then the
effective angle defining the shadow boundary is less (in absolute
value) then the incident angle $\t$, and if $\pi/2<|\t|<\pi$ that
angle is greater than $\t$. In both cases there is a rotation of
the shadow boundary towards the negative $x$ half-axis as compared
to the stationary case. Thus wave lit and shadow domains are
modified with respect to stationary case. With changing $t$ this
line is shifted parallel in the $x$-direction.

\begin{figure}[h]
\begin{center}
\includegraphics[width=.7\textwidth]{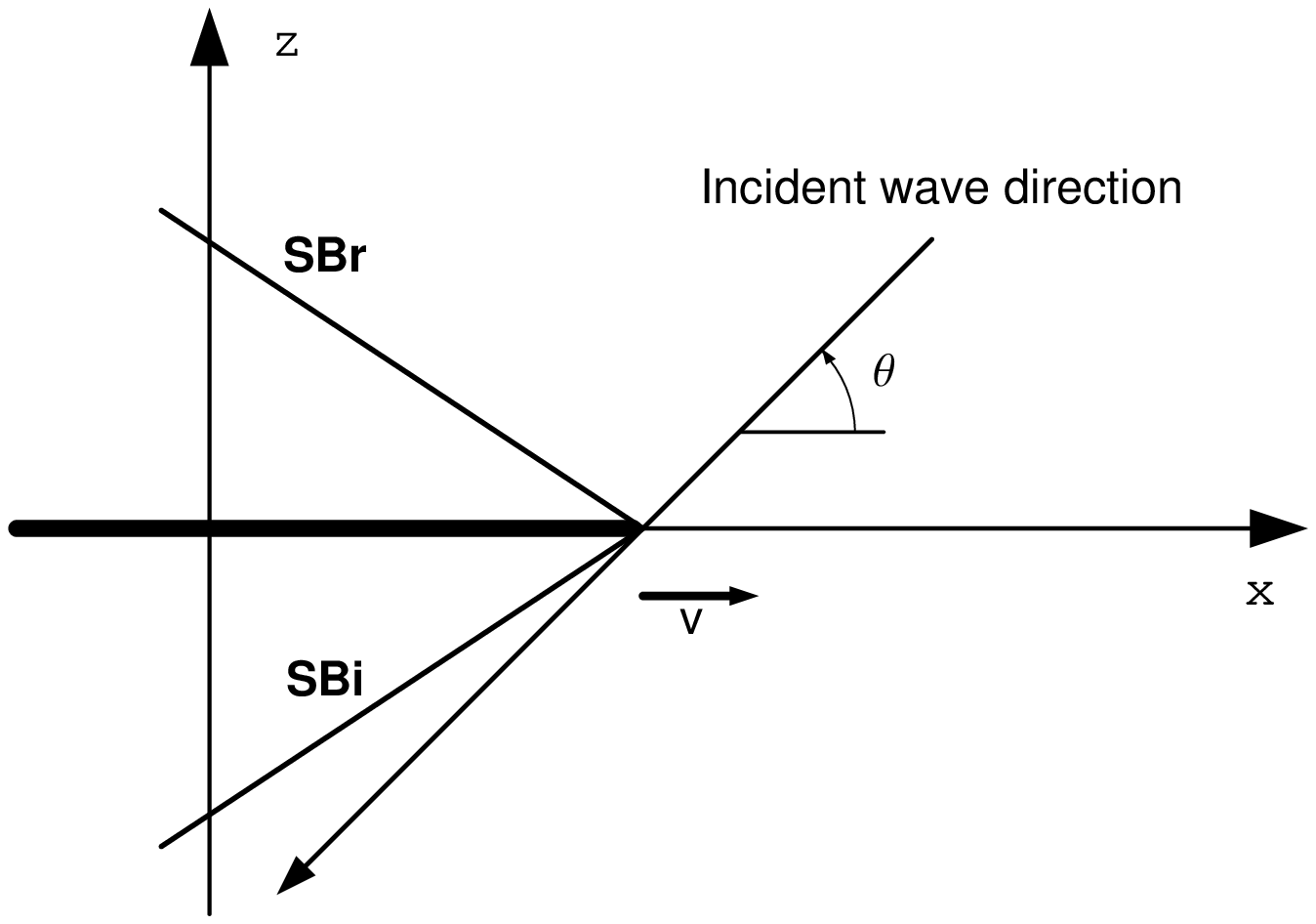}
\end{center}
\end{figure}
\begin{center}\parbox{.8\textwidth}{Fig.1 \textit{Rotation of the shadow
boundaries of the incident (SBi) and the reflected (SBr) waves for
a moving half-plane.}}
\end{center}

\vspace{2ex} In the moving frame the diffracted wave can be
physically interpreted as a cylindrical wave emanated from the
half-plane edge. Its phase function is given by $k\p \r\p - k\p
ct\p - \pi/4$. Levels of constant phase in $(x\p,z\p)$ space are
coaxial, circular cylinders centered at $\r\p=0$. In the
laboratory frame it is given by
\begin{equation}\label{e41}
k\g^2(1+\b\cos{\t})\left[\sqrt{(x-\b ct)^2 + (z/\g)^2}+\b
x-ct\right] - \pi/4,
\end{equation}
(see (\ref{e37})).

Equiphase surfaces in the laboratory frame are described by the
equation
\begin{equation}\label{e42}
\sqrt{(x-\b ct)^2+(z/\g)^2}+\b x-ct = C, \qquad C=\mbox{const}.
\end{equation}
It can be readily seen that for fixed $C$ extremal values of $z$
occur at $x=\b\g^2 C$. Now (\ref{e42}) can be transformed into the
form
\begin{equation}\label{e43}
(x+\b\g^2 C)^2+z^2=(ct+\g^2 C)^2,
\end{equation}
describing a circle of radius $ct+\g^2 C$, centered at
$(-\b\g^2C,0)$. The distance of the half-plane edge from the
circle center is $\b (ct+\g^2 C)$, i.e.\ it is $\b$ times smaller
than its radius. Thus for $\b<1$ the edge is located always inside
the circle, and it enters the circle if $\b=1$. As $C$ increases,
the circle centers are shifted along the $x$-axis backwards with
respect to the direction of edge motion (see Fig.2)). This
equation was first obtained in a different, but equivalent form in
\cite{lm;67}, where diffraction on a moving thin cylinder was
examined.

\begin{figure}[h]
\begin{center}
\includegraphics[width=.7\textwidth]{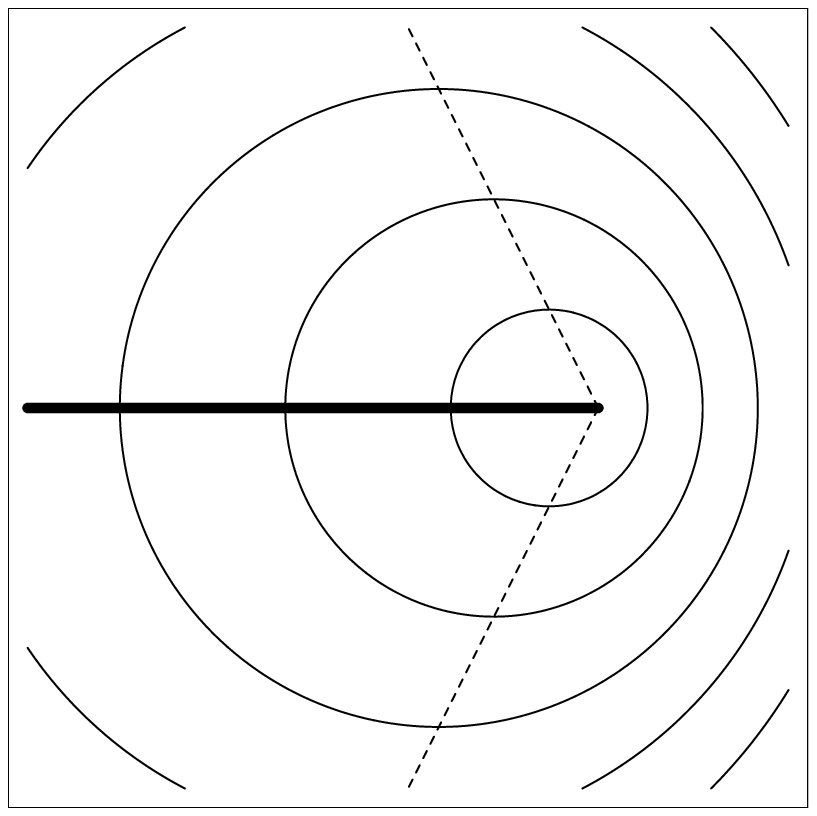}
\end{center}
\end{figure}
\begin{center}\parbox{.8\textwidth}
{Fig.2 \textit{Equiphase lines in the $(x,z)$ plane. The dashed
rays are drawn through the maximal and minimal values of $z$ on
them.}}
\end{center}

\newpage
The loci of points corresponding to extremal values of $z$ for
different $C$ are two semi-infinite straight lines
\begin{equation}\label{e44}
\frac{|z|}{x-vt}=-\frac{1}{\b}\: ,
\end{equation}
(in primed coordinates they are given by $\b+\cos{\f\p}=0$). It is
easy to see that along these lines $h_3=0$, and hence the
$z$-components of the diffracted field also vanish on them.
Additionally, $z$-component of phase velocity attains its maximal
value $c$ on these lines.

Singularities of the diffracted wave occur at all point where the
denominator in $g^\pm$ is zero, i.e.\ where
\begin{equation}\label{e45}
(1+\b\cos{\t})(x-\b ct)+(\cos{\t}+\b)r=0.
\end{equation}
It then follows that the singularities appear on the straight
lines
\begin{equation}\label{e46}
\frac{|z|}{x-\b ct}=\frac{\sin\t}{\cos{\t}+\b}\: , \qquad
\mbox{sign}(x-vt)=-\mbox{sign}(\cos{\t}+\b).
\end{equation}
This fact was observed in \cite{gt;68} in the study of moving
wedge diffraction. These lines are now recognized as coinciding
with the shadow boundaries (\ref{e40}).

The fact that the equiphase surfaces in the diffracted wave are
circles in the $(x,z)$ plane is obvious --- in a homogeneous
medium EM signals excited from a point source propagate with the
same velocity in all directions. Moreover, disturbances originated
at the half-plane edge lag behind the edge as it is moving
forward. This explains both the rotation of the shadow boundaries,
as well as the shift of the equiphase surface centers backwards
with respect to the edge. Above phenomena are similar to those
which can be observed on water surface when disturbances are
evoked on one side of a moving ship and at its bow.

\section{Conclusions}
In this work we examined 2D problem of EM plane wave diffraction
on a moving half-plane. The classical Sommerfeld solution for a
stationary problem and Lorentz transformation were used to obtain
the solution in the laboratory frame. The field obtained is no
longer time-harmonic, as it is in the stationary case. We expanded
the exact solution asymptotically, and interpreted the result in
physical terms. It appeared that for the half-plane displacement
considered in this work the geometrical optics field has a similar
form as in the stationary problem, except that the illuminated
region of the incident wave and the shadow region in the case of
the reflected wave increase in comparison to the stationary case.
It was also shown that the surfaces of constant phase of the
diffracted wave are circular cylinders, but they are not
concentric, as they are in the stationary case. Even though these
phenomena are most pronounced when $\b$ is close to $1$, the
results obtained are valid for arbitrarily $\b$. If $\b$ is
sufficiently small, these results can be simplified by replacing
$\g$ with its expansion for small argument, or even by putting
$\g\approx 1$.

%
%\newpage\centerline{\textbf{\large Figure captions}}
%
%\vspace{4ex} Fig.1 Rotation of the shadow boundaries of the
%incident (SBi) and the reflected (SBr) waves for a moving
%half-plane.
%
%\vspace{2ex} Fig.2 Equiphase lines in the $(x,z)$ plane. The
%dashed rays are drawn through the maximal and minimal values of
%$z$ on them.

\end{document}